**Title: Not Yet for Us: The Nascent Black Hole**

**Author: Martin Hill**

**Abstract**

In the second half of the last century the 'frozen star' or 'collapsed star' models of black hole formation were largely abandoned. The result is that later models appear to either assume that black hole event horizons already exist (such as in Hawking's 1975 paper on radiation from black holes), bypass the relative frames altogether (such as in Mirabel's 2017 paper on the formation of stellar black holes), or use coordinate systems that essentially ignore the remote viewer's point of view (Kruskal 1960).

This paper attempts to re-establish the concept of a 'nascent black hole' as the correct approach for modelling black holes from remote reference frames. It uses, and only needs to use, Schwarzschild metrics and presents some example scenarios to demonstrate the concepts through worked examples. Alternatives such as Eddington-Finkelstein coordinates and Penrose's local collapsing frames are not disputed; the Schwarzschild metrics must still be valid outside the event horizon and this paper is largely concerned with 'remote viewers' for various meanings of 'remote'.

For most astrophysical or cosmological questions the difference between 'actual' and 'nascent' black holes is largely irrelevant as they behave similarly for most practical purposes. However the difference becomes critical for effects that require an event horizon to interact with the remote viewer, such as Hawking-Zel'dovich radiation and the related information problems outlined by Susskind. That is: if there are no possible observable paths from a nascent event horizon to any remote viewer then there also cannot be any evaporation to the remote viewer.

1. **Background and History**

In the second half of the last century the 'frozen star' or 'collapsing star' models of black hole formation were largely abandoned (Ruffini and Wheeler 1971; Thorne 1994; Thorne and Price 2009). This appears to have been partly due to a switch of viewpoint to the collapsing material rather than the remote viewer, which does not 'freeze' but continues to collapse in its own time. New mathematical models (Kruskal 1960) provided an alternative to the Schwarzschild metrics which had restricted what could be understood beyond the event horizon due to its coordinate singularity at that point.

However the asymptotically-remote and asymptotically-close-to-horizon views do not appear to have been properly resolved. Schwarzschild metrics are accepted as being valid outside the event horizon (Thorne 1994 p2-3) yet the consequences seem to be discarded when considering events close to the horizon. For example, observations by remote viewers of events near the horizon are described as a 'distorted picture of "Reality"' (Thorne and Price 2009, 27). Hawking declares the singularity as 'fictitious' (Hawking 1975) which may be correct for the coordinates but not for the effects of being there. Bunn describes effects as 'really just an optical illusion' and that Schwarzschild coordinates give a 'highly distorted *view* of what's going on near the horizon' (Bunn 1995, my emphasis) although he does also consider the effects of re-uniting frames. Some other contradictions may be due to casual language; Kip Thorne's entertaining prologue in 'Black Holes & Timewarps: Einstein's Outrageous Legacy'includes a remote viewer seeing Arnold the Android fall *into* the event horizon even though the 'forever-doubling signals take forever long to climb out of the hole's gravitational grip' (Thorne 1994)

In many cases the frames do not need to be resolved. For example the nature of materials and their behaviour when collapsing into black hole singularities do not need to refer to a remote reference frame (Chandrasekhar 1984; Mirabel 2017). The large-scale effects of a massive collapsing body, such as gravitational frame dragging, do not need an extant event horizon to be valid. However some effects, such as Hawking-Zel'dovich Radiation, require a causal relationship between an event horizon and a remote viewer in order to occur.

## 2. Introduction & Assumptions

This paper attempts to clarify some concepts in the relationships between the asymptotically-remote and asymptotically-close-to-horizon reference frames in order to show that effects that require an event horizon cannot affect remote viewers.

It takes three approaches in two scenarios, each considering several reference frames. The two scenarios are (1) bodies near and far from an event horizon and (2) a forming event horizon. The three approaches are (a) conceptual (b) Schwarzschild metric worked examples and (c) space-time diagramming.

Only events and concepts in space-time outside and excluding the nascent event horizon of a non-rotating black hole are given. Schwarzschild geometry is therefore sufficient. Other coordinate frames may handle situations around and below the event horizon without the inconvenience of the Schwarzschild singularity but they must nevertheless map to Schwarzschild outside the horizon. That is: any explanation in Schwarzschild geometry, as long as it is outside or before a formed event horizon, is valid for all other geometries too.

Only radial paths need to be considered, so the Schwarzschild metric (eg Longair 1984, sec. 14.6; Foster and Nightingale 1976, sec. 4.1) coordinate mapping can be simplified to:

$$d\tau^2 = \left(1 - \frac{2GM}{c^2 r}\right) dt^2 - \frac{dr^2}{c^2 \left(1 - \frac{2GM}{c^2 r}\right)} \tag{1}$$

This is further simplified by considering a particular nascent black hole so the Schwarzschild radius can be substituted as $r_h$, ie

$$r_h = \frac{2GM}{c^2} \tag{2}$$

and:

$$d\tau^2 = \left(1 - \frac{r_h}{r}\right) dt^2 - \frac{dr^2}{c^2 \left(1 - \frac{r_h}{r}\right)} \tag{3}$$

For most of the examples the frames will be static with respect to each other, so only the left hand term will be required.

### 3. Back to Basics: Four People with Clocks

This section introduces four conveniently labelled frames of reference to provide the different relative points of view. These frames are represented as people that were all collocated at some earlier point, each with a highly luminous clock synchronized with all the others.

In the first scenario they are arranged around an existing super-massive black hole with a mass of around $10^{10}$ Solar masses (by using such a mass distractions such as tidal forces can be discarded). The frames are populated as follows (see Figure 1):

- Remote Robin, positioned 'far' away in asymptotically flat space
- Static Sam, positioned close enough to the event horizon to be affected by it
- Yoyo Jo who travels between Remote Robin and Static Sam.
- Free-falling Freddy who falls towards the event horizon

Remote Robin is positioned far enough away to be considered in 'flat space'. A rope dangles from Remote Robin to near the black hole's event horizon, its top end kept 'fixed' in place near Robin. (The rope is only a 'conceptual handrail' for movement up and down the gravity well). Static Sam is hanging on to the rope nearer to the event horizon. Yoyo Jo travels up and down the rope between Remote Robin and Static Sam. Free Falling Freddy travels down the rope and then lets go to free-fall into the hole.

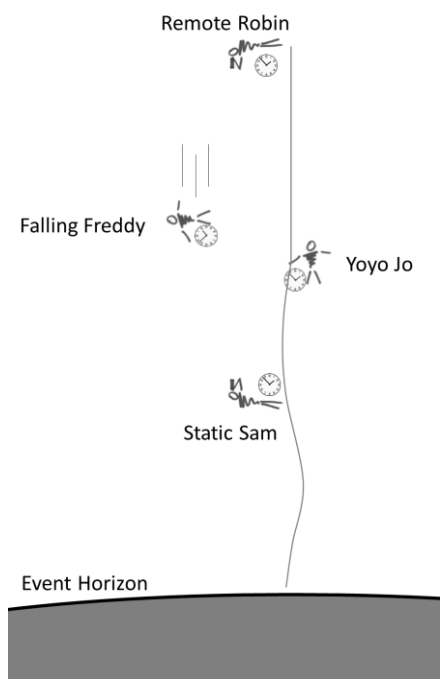

**Figure 1 - Four Frames**

All parties can communicate the frequency and brightness of what they emit and observe according to their own clocks, independent of their position or clock speed. For example each can declare "My clock is emitting 1,000 photons per tick with a frequency of 500 THz" and this *information* will be received as it was given no matter the distortions in the means of communication.

As there is no change in distance between Sam and Robin, Static Sam's clock observed by Remote Robin appears to run slow by the following component of the Schwarzschild equation in (3):

$$d\tau^2 = \left(1 - \frac{r_h}{r}\right) dt^2 \qquad (4)$$

To begin with Static Sam is positioned so that Sam's clock appears to tick half as often as Remote Robin's which (see Appendix – Workings 1) gives Static Sam a position about 1/3 further out from the centre than the nominal event horizon. A mass of $10^{10}$ solar masses has an event horizon at $r_h$ of about $3 \times 10^{11}$ km (See Appendix – Workings 2), so Static Sam is about $10^{11}$ km from that. It is not important in this case which distances are measured with what ruler.

The difference in clock speeds means that Remote Robin will measure light from Static Sam's clock as 'darker' than Sam declares. If Sam declares that 1,000 photons per tick of Sam's clock are emitted towards Robin, then Robin will see only 500 photons per tick of Robin's clock as the natural consequence of Robin seeing Sam's clock run at half speed. This seems to be reasonably uncontroversial.

However agreement of Static Sam's view appears to be less well settled. This is partly because convention usually considers only a Remote Robin and a Free Falling Freddy, and partly because it appears tempting to use universal or absolute clocks and declare differences as only 'apparent' as outlined in Section 1.

In this thought experiment Yoyo Jo is introduced to carry a frame of reference between Static Sam and Remote Robin in order to demonstrate that the dilation effects on Remote Robin's view of Static Sam must be balanced by the opposite effect on Static Sam's view of Remote Robin.

Jo starts with a common frame to Robin and travels down the rope at a constant speed to Sam, then up it to Robin again, observing Robin's clock during the whole journey. This is shown as a space-time diagram in Figure 2 with horizontal bars that represent clock ticks as observed by Jo as Jo travels up and down the rope. Note that in this space-time diagram, the 'time' axis is with respect to Yoyo Jo's frame of reference; it is linear in Jo's 'proper' or local time.

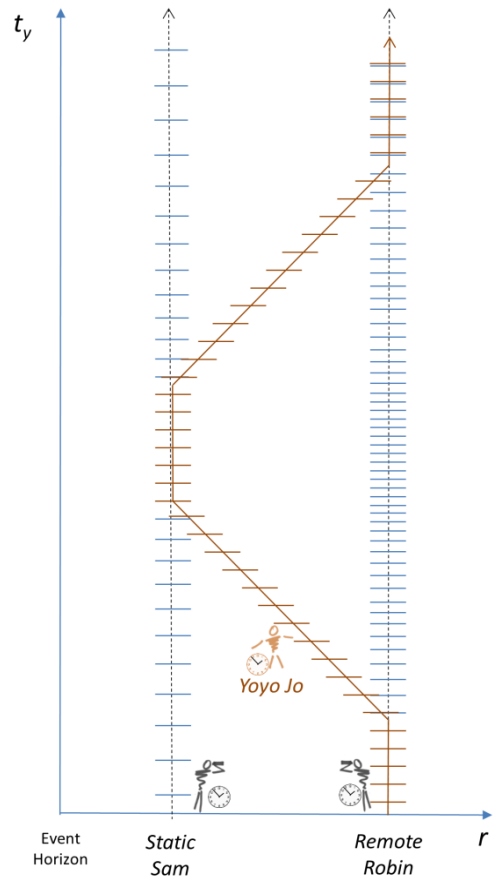

**Figure 2 - Yoyo Jo**
**(time axis w.r.t. Yoyo Jo frame)**

Robin will see Jo's clock tick slow as it travels away and then speed to Robin's local rate as it returns. (There are also apparent changes to clock speeds due to relative acceleration and velocities but these are not relevant here).

By the time Yoyo Jo returns to Remote Robin, Jo's clock will not be as advanced as Robin's. Robin will have seen Jo's clock run slow while Jo was visiting Sam, and correspondingly, Yoyo Jo will have observed Robin's clock tick more times than Jo's local one between departure and return. Jo must therefore also have observed a higher brightness (photons received per Jo's *local* tick) of Robin's clock than it emitted.

This re-uniting of frames shows that conceptually the view of Robin from Jo deep in the well *must* correspond inversely to the view of Jo from Robin 'outside' the well. This 'real' difference in elapsed time has been confirmed experimentally using nuclear resonance (Pound and Rebka 1959) and in the lab using optical clocks (Chou et al. 2010) amongst other confirmations of relativity (Vessot 1979; Nordtvedt 1982).

As for Jo, so for Sam. If Static Sam is positioned closer to the event horizon and Yoyo Jo visits Sam for longer periods, the difference between Yoyo Jo and Remote Robin's re-united clocks increases. Robin will observe Sam and Jo's clocks as slower and darker. Correspondingly Robin's clock must appear, to Sam and visitor Jo, to become faster and brighter. As for Remote Robin so for all else in the remote universe that Sam and visitor Jo observes; the remote universe appears to run faster and so brighter.

This discrepancy is important to avoid the tempting misconception of a universal or absolute time. Stannard is not alone, for example, in saying that while a remote observer might be unable to see a spaceship fall through the event horizon "the craft itself has *actually* passed through that region quite quickly" (Stannard 2008, 86, my emphasis). Similarly Kip Thorne describes Android Arnold who "flew through the horizon, moving at the speed of light, *many minutes ago*" and the impression of Arnold lingering near the event horizon are merely "relics from the past." (Thorne 1994, 33, my emphasis). Bunn says that the best way of thinking about is as "just an optical illusion" and that Schwarzschild coordinates give a "highly distorted *view* of what's going on near the horizon" (Bunn 1995, my emphasis).

A worked example may be useful. A $10^{10}$ solar mass black hole has an event horizon $r_h$ of about $3 \times 10^{13} m$ (see Appendix, Workings 2). Sam descends to a million kilometres from the event horizon ($r = r_h + 10^9 m$), which gives a proper time ratio of (approximately):

$$d\tau = \left(1 - \frac{r_h}{r}\right) = \left(1 - \frac{3 \times 10^{13}}{3.0001 \times 10^{13}}\right)^{1/2} dt = 0.006 \, dt \qquad (5)$$

Yoyo Jo descends to Static Sam's position and, after a year according to the Sam/Jo clocks, ascends back to Remote Robin. Robin's clock will have advanced by over 166 years compared to Jo's (not including the time it would take for Yoyo Jo to descend and ascend). During that year-long visit according to the Sam/Jo clocks, Sam and Jo will have watched Remote Robin's clock advance 166 years – and received radiation from the remote universe emitted over those 166 years.

If Sam is positioned at slightly more than one meter from this event horizon, Robin's clock will advance by $10^7$ years for each of Jo's years spent with Sam. That means that Jo and Sam receive $10^7$ years' worth of remote universe radiation for each of their local years. The universe, to them, is running very fast and very bright.

As Sam's position 'tends to' the event horizon, so the difference in clocks between Jo and Robin 'tends to' infinity: Robin sees Sam's clock slow to a stop - and Sam sees Robin's clock speed to infinitely fast.

These are therefore not just 'relics' of an event that has occurred to Sam in Robin's past; they are features of the difference in time experienced due to the differences in position in the gravitational well. The Schwarzschild metrics do not just 'distort the view' of what's going on near the event horizon, they *describe* the distortion of space-time and their consequences.

Free Falling Freddy is not required to illustrate these concepts but many texts compare a Freddy travelling over the event horizon with Remote Robin. In these cases Freddy's velocity rapidly becomes relativistic, and this complicates the picture as there are now a number of contributing modifiers. Due to the gravity well Robin would see Freddy's clock run slower and Freddy would see Robin's clock run faster. Their difference in velocity means each see the other's clock run slower. The space around Freddy's frame is increasingly curved, but Freddy is in a locally inertial frame that is nevertheless accelerating with respect to Remote Robin.

However, while interesting, Freddy's view does not supersede Yoyo Jo's view. Consider if Freddy lets go of the rope a thousand kilometres ($10^6$ m) away from the event horizon. Freddy will initially accelerate at about $10^7$ m/s$^2$ and so will reach the event horizon in a fraction of a second at a non-relativistic relative velocity. Falling Freddy's experiences are therefore similar to Static Sam's when Sam was positioned closer to the event horizon. Freddy will see the rest of the universe, including Robin's clock, appear faster and brighter. As with Sam, as Freddy's position 'tends to' the event horizon, the remote clocks 'tend to' infinite speed according to Freddy.

The use of a Falling Freddy in texts appears only to introduce the misconception that since a short, finite time has passed for Freddy then Freddy has 'actually' passed the event horizon according to Robin. This 'universal time' misconception allows events that follow the crossing to then affect the remote universe (see Figure 3).

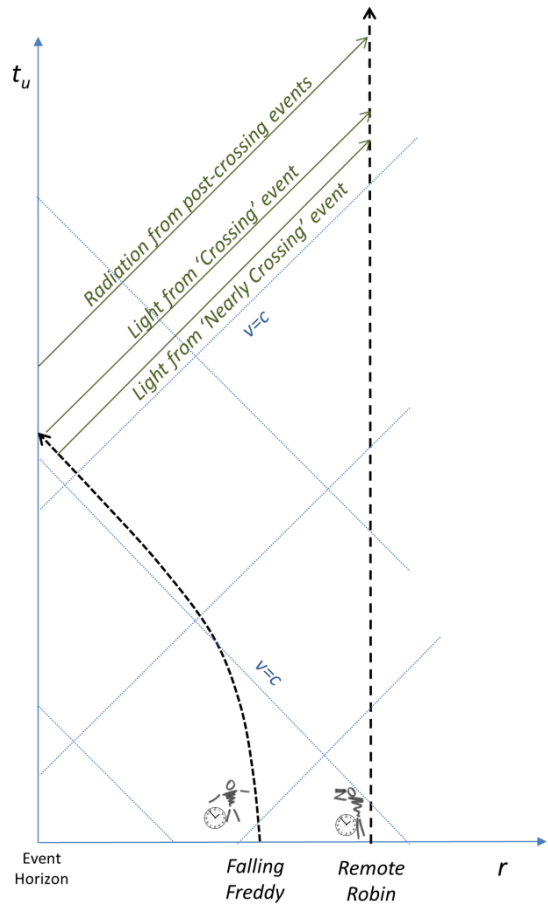

Figure 3 - Misconception: Freddy Falls in "universal frame"

As we saw above however the two views – Freddy falls in finite falling time according to Freddy and infinite time according to Remote Robin – are both 'true' and a natural result of the space-time curvature.

In any case events that occur to Freddy at or after crossing the event horizon cannot reach Remote Robin (see Figure 4). Emissions from Free-Falling Freddy as Freddy 'reaches' the event horizon have the most space-like travel possible. There are no paths from Free-Falling Freddy that can reach Remote Robin faster than the light from Freddy asymptotically near the event horizon. As long as Remote Robin sees – or could see - a highly-red-shifted Freddy-clock, Robin cannot then also see the consequences of Freddy's crossing. This includes changes to the size of the event horizon or changes to the mass of the black hole.

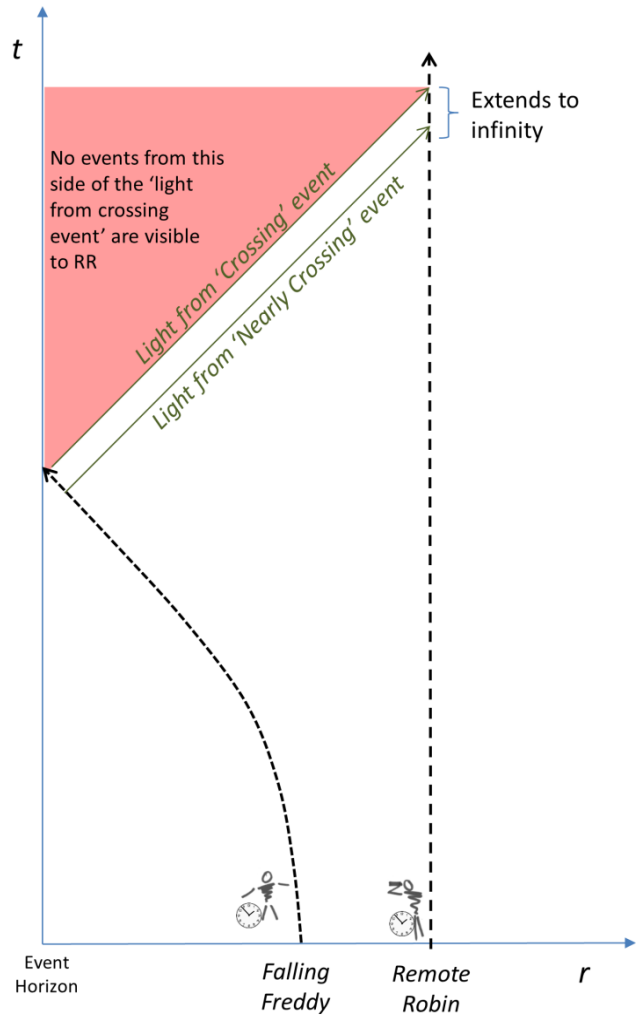

**Figure 4 - Invisible Freddy with 'nominal' time axis**

It is true that Freddy may only count a few clock ticks while falling to and through the event horizon but it is not useful to then map this onto the equivalent number of Robin clock ticks. Freddy's time no longer corresponds in any meaningful way to Robin's time. As Freddy or Sam or Jo 'reach' the horizon their view or Robin and the rest of the universe shows it increasingly fast until it comes to an end; space around them might be said to become 'undefined.'

Some texts (eg Stannard 2008, p88) claim that 'nothing special' happens at the event horizon, possibly to make the point that there is no 'thing' or surface at that point. However there are noticeable effects in the form of lensing, time dilation and blue-shift (Nemiroff 1993). The asymptotically increasing blue-shift, and condensing of the remote universe's emissions into shorter and shorter local time, implies that any Falling Freddy approaching the event horizon will be passing through a firestorm of accumulated radiation.

## 4. Not even a Brown Crusty Ring: The Nascent Black Hole

The above section is mainly intended to revisit the relevant frames and answer some possibly muddying misconceptions. It argues that given a black hole with an existing event horizon then any events that occur to a body at or after reaching the event horizon have no time-like path to, and so cannot affect, any remote observer.

This section uses a similar approach to the formation of event horizons in the first place. The time according to the collapsing body will be finite, but according to the remote observer the collapse tends to an infinite time. This is not incompatible and is simply a feature of the relativistic difference in local times. Because the formation of the event horizon is not visible to the remote observer therefore no subsequent events or effects can be.

Consider for example the formation of a $10^{10}$ solar mass black hole. By using such a large mass (about $2 \times 10^{40} kg$) we can consider a formation without relativistic motion or exotic matter; it 'merely' requires a large, cool dust cloud to start densely enough. Such a mass has a Schwarzschild radius of about $3 \times 10^{13} m$ so if we start this cloud at about twice the Schwarzschild volume it will have a density of less than 1 kg m$^{-3}$ (See Appendix Workings 3).

As before Remote Robin is far enough away to be in asymptotically flat space, and 'Surface Sam' is floating in the outermost layers of the dust cloud.

Schwarzschild metrics still apply. As the cloud contracts and takes up a smaller volume Remote Robin sees Surface Sam's clock slow and Sam sees Robin's speed up. When the outer limit of the cloud has a radius of $4/3 r_h$ (see Appendix Workings 1) Surface Sam's clock will be half the speed of Robin's. Yoyo Jo can travel up and down a rope between them to marry up their frames in just the same way as before.

As the cloud contracts further and the radius of the cloud reaches just over 1 meter more than $r_h$, Sam's clock ticks once for every $10^7$ ticks of Remote Robin's. Robin will see a slow, dark and reddened Sam-clock, Sam and cloud. Sam sees a fast, bright and blue remote Robin-clock, Robin and universe. As the cloud condenses further Robin sees it become darker and redder and slower until it become effectively invisible. Sam sees Robin's clock speed to infinity along with the motion and blue-shifting of the rest of the universe, lensed into a smaller and smaller angle of the sky.

Similarly a collapsing star would also become darker, browner, and slower to Robin as it approaches the Schwarzschild limit. A possible misconception is that the 'frozen star' is a dark, red 'crust' of matter accumulating around an event horizon at the Schwarzschild limit. This is unnecessary; the difference in space curvature causes the time dilation and so apparent slowing rather than something 'at' the Schwarzschild radius. There is also no need for a supporting physical mechanism. The star still collapses in its own finite local time; the local 'co-moving' collapsing frame and asymptotically flat remote frame are running at different rates.

5. **Conclusion**

For most cosmological and astrophysical situations a nascent black hole will behave similarly to a black hole with an existing event horizon: its gravity, charge and rotation will affect surrounding material in the same way. A nascent black hole may also be so very dark and red-shifted that it is effectively black. The dilation effects tend to occur very close to the Schwarzschild radius so the size of the nascent black hole will be very similar to an 'actual' black hole.

However events that require an event horizon to be 'available' on a time-like path to a remote viewer cannot have an effect. Changes to event horizons or transitions of event horizons cannot affect the remote frame if the event horizon is not yet visible to that frame.

In principle re-establishing a variant of the 'frozen star' as a nascent black hole solves a number of puzzles by pushing them to a different universe. For example the hairiness or otherwise of event horizons is moot if they have no effecting presence on the remote universe.

## 6. Limitations

If the plank length is a lower limit to meaningful distance then it may be that the mathematical asymptotic limits to particular distances are not suitable. In this paper I have used the Schwarzschild radius as the asymptotic limit for external bodies approaching the event horizon and the formation of an event horizon. It might be that once material has collapsed to 'close enough' to that radius they have effectively 'reached' the Schwarzschild radius. Time dilation at a plank length (approx. $10^{-35}$ m) from the event horizon is about $10^{17}$ and the universe is about that many seconds old; that is, such black holes would be about one second old in their local time.

In principle Hawking-Zel'dovich radiation does not require an event horizon; it may be possible for a nascent black hole or any other sufficiently dense body to effectively radiate where the gravity gradient is sufficiently curved that pair-particles effectively separate. This suggests the energy 'lost' would be from the gradient rather than the conceptual 'surface' of an event horizon, and would similarly suggest that 'hairiness' would be a general property of curved space.

**Appendix – calculations & working**

<u>Working 1: Sam at half clock speed of Robin</u>

$$d\tau = 1/2\, dt$$

Substituting into equation (4) $d\tau^2 = \left(1 - \frac{r_h}{r}\right) dt^2$

$\Rightarrow \frac{1}{2} = \sqrt{\left(1 - \frac{r_h}{r}\right)}$

$\Rightarrow \frac{1}{4} = \left(1 - \frac{r_h}{r}\right)$

$\Rightarrow \frac{3}{4} = \frac{r_h}{r}$

$\Rightarrow r = \frac{4}{3} r_h$

<u>Working 2: Schwarzschild radius for $10^{10}$ Solar masses</u>

$$M_{solar} = 2 \times 10^{30} kg$$

$$r_h = \frac{2GM}{c^2}$$

$$= \frac{2 \times 7 \times 10^{-11} \times 2 \times 10^{30} \times 10^{10}}{(3 \times 10^8)^2} \frac{kg\, m^3}{kg\, s^2 \left(\frac{m}{s}\right)^2}$$

$$= \frac{28 \times 10^{29}}{9 \times 10^{16}} m$$

$$= \sim 3 \times 10^{13} m$$

<u>Working 3: Density of nascent black hole collapsing cloud</u>

$10^{10}$ Solar masses = ~ $2 \times 10^{40} kg$

Schwarzschild radius of $10^{10}$ Solar masses as above = ~$3 \times 10^{13} m$

(For comparison 1 AU = $1.5 \times 10^{11} m$)

Volume = 4/3 π r³ = ~ 100 x $10^{39}$ m³ = $10^{41}$ m³

So density of a cloud at about twice the volume would be about 1 kg/10 m^3